\input{aipcheck}

\documentclass[
    ,final            
,numberedheadings 
  ]
  {aipproc}

\layoutstyle{6x9}

\def\rn{}
\def\nn#1 #2{#2. #1}				
\def\nnn#1 #2 #3{#2. #3. #1}			
\def\nnnn#1 #2 #3 #4{#2. #3. #4 #1}		
\def\nnnnn#1 #2 #3 #4 #5{#2. #3. #4 #5. #1}	
\def\dualand{ and\hbox{ }}				
\def\multiand{, and\hbox{ }}				
\def\rf#1;#2;#3;#4;#5 {{\frenchspacing\par\rn#1, #3 {\bf #4}, #5 (#2). \par}}
\def\rrf#1;#2;#3;#4;#5 {{\frenchspacing\rn#1, #3 {\bf #4}, #5 (#2);~}}
\def\rrrf#1;#2;#3;#4;#5 {{\frenchspacing\rn#1, #3 {\bf #4}, #5 (#2).}}
\def\rg#1;#2;#3;#4;#5;#6 {{\frenchspacing\par\rn#1, #3 {\bf #4}, #5 (#2). \par}}
\def\rfbook#1;#2;#3;#4;#5 {{\frenchspacing\par\rn#1, {\it #3} (#5, #4, #2).\par}}
\def\rfprep#1;#2;#3 {{\par\frenchspacing\rn#1, #3 (#2).\par}}
\def\rrfprep#1;#2;#3 {{\frenchspacing\rn#1, #3 (#2);~}}
\def\rrrfprep#1;#2;#3 {{\frenchspacing\rn#1, #3 (#2).}}
\def\rfproc#1;#2;#3;#4;#5;#6 {{\frenchspacing\par\rn#1 #2, in {\it #3}, ed. #4 (#5: #6)\par}}
\def\rfprocp#1;#2;#3;#4;#5;#6;#7 {{\frenchspacing\par\rn#1 #2, in {\it #3}, ed. #4 (#5: #6), p#7\par}}

\def\rg#1;#2;#3;#4;#5;#6 {\par\rn#1 #2, {\it #3}, {\bf #4}, #5 (``#6'') \par}
\def\rf#1;#2;#3;#4;#5 {\par\rn#1, {\it #3}, {\bf #4}, #5 (#2)\par}
\def\rfbook#1;#2;#3;#4;#5 {{\frenchspacing\par\rn#1, {\it #3} (#4: #5, #2)\par}}
\def\rfproc#1;#2;#3;#4;#5;#6 {{\frenchspacing\par\rn#1 #2, in {\it #3}, ed. #4 (#5: #6)\par}}
\def\rfprocp#1;#2;#3;#4;#5;#6;#7 {{\frenchspacing\par\rn#1 #2, in {\it #3}, ed. #4 (#5: #6), p#7\par}}
\def\rfprep#1;#2;#3  {{\par\rn#1, #3, #2\par}}
\def\rfprepp#1;#2;#3 {{\par\rn#1 #2, #3\par}}

\def\etal{{\frenchspacing\it et al.}}

\def\eg{{\frenchspacing\it e.g.}}

\def\bfk{\mbox{\bf k}}

\def\bfx{\mbox{\bf x}}

\newcommand{\be}{\begin{equation}}
\newcommand{\ee}{\end{equation}}
\newcommand{\ba}{\begin{eqnarray}}
\newcommand{\ea}{\end{eqnarray}}

\def\fun#1#2{\lower3.6pt\vbox{\baselineskip0pt\lineskip.9pt
        \ialign{$\mathsurround=0pt#1\hfill##\hfil$\crcr#2\crcr\sim\crcr}}}
        
\def\beq#1{\begin{equation}\label{#1}}
\def\eeq{\end{equation}}
\def\beqa#1{\begin{eqnarray}\label{#1}}
\def\eeqa{\end{eqnarray}}

\def\spose#1{\hbox to 0pt{#1\hss}}
\def\simlt{\mathrel{\spose{\lower 3pt\hbox{$\mathchar"218$}}
     \raise 2.0pt\hbox{$\mathchar"13C$}}}
\def\simgt{\mathrel{\spose{\lower 3pt\hbox{$\mathchar"218$}}
     \raise 2.0pt\hbox{$\mathchar"13E$}}}
\def\simpropto{\mathrel{\spose{\lower 3pt\hbox{$\mathchar"218$}}
     \raise 2.0pt\hbox{$\propto$}}}

\def\ed{\end{document}}

\def\Om{\Omega_m}

\def\beq#1{\begin{equation}\label{#1}}
\def\eeq{\end{equation}}
\def\beqa#1{\begin{eqnarray}\label{#1}}
\def\eeqa{\end{eqnarray}}

\newcommand{\Nifs}{\ensuremath{^{56}\mathrm{Ni}}}
\newcommand{\msun}{\ensuremath{M_\odot}}

\newcommand{\Hunit}{\,{\rm km}\,{\rm s}^{-1}\,{\rm Mpc}^{-1}}

\begin{document}
\input{epsf.sty}
\input{psfig.sty}

\title{Observational Probes of Dark Energy}


\author{Yun Wang}{
  address={Homer L. Dodge Department of Physics \& Astronomy, Univ. of Oklahoma,
                 440 W Brooks St., Norman, OK 73019; wang@nhn.ou.edu}
}

\begin{abstract}
The cause for the observed acceleration in the expansion of the universe is unknown, and referred to as ``dark energy'' for convenience. Dark energy could be an unknown energy component, or a modification of Einstein's general relativity. This dictates the measurements that are optimal in unveiling the nature of dark energy: the cosmic expansion history, and the growth history of cosmic large scale structure. I will examine Type Ia supernovae and galaxy clustering as dark energy probes, and discuss recent results and future prospects.
\end{abstract}

\maketitle


\section{Introduction}

Solving the mystery of the observed cosmic acceleration \cite{de}
is one of the most important challenges in cosmology today. Current observational 
data are not sufficient for differentiating two likely explanations for the observed
cosmic acceleration: an unknown energy component (dark energy, \eg, \cite{quintessence}), 
and the modification of general relativity (modified gravity, \eg, 
\cite{modifiedgravity,DGPmodel}). 
Refs.\cite{reviews,Wang_book} contains reviews with more complete 
lists of references of theoretical models.

The evidence for cosmic acceleration has strengthened over time.
The expansion history of the universe is described by the
Hubble parameter, $H(t)=(d\ln a/dt)=\dot a/a$, where $a(t)$ is the
cosmic scale factor, and $t$ is cosmic time.
The cosmological redshift, $z \equiv 1/a(t) -1$, is usually used
as the indicator for cosmic time, because it can be measured
for a given astrophysical object.
Fig.1 (left panel) shows the Hubble parameter $H(z)$, as well as
$\dot a$, measured from current observational data \cite{Wang11}.
Fig.1 (right panel) shows the corresponding dark energy density
function $\rho_X(z)$ \cite{Wang11}.
A cosmological constant (and no modification of gravity)
continues to be consistent with data, but the uncertainties are large 
(see Fig.1, right panel). Given our ignorance of the true nature of
dark energy, and the theoretical difficulties of explaining a tiny
but non-vanishing cosmological constant using known physics, we
need to be open minded in exploring and constraining alternative
explanations.

\begin{figure} 
\psfig{file=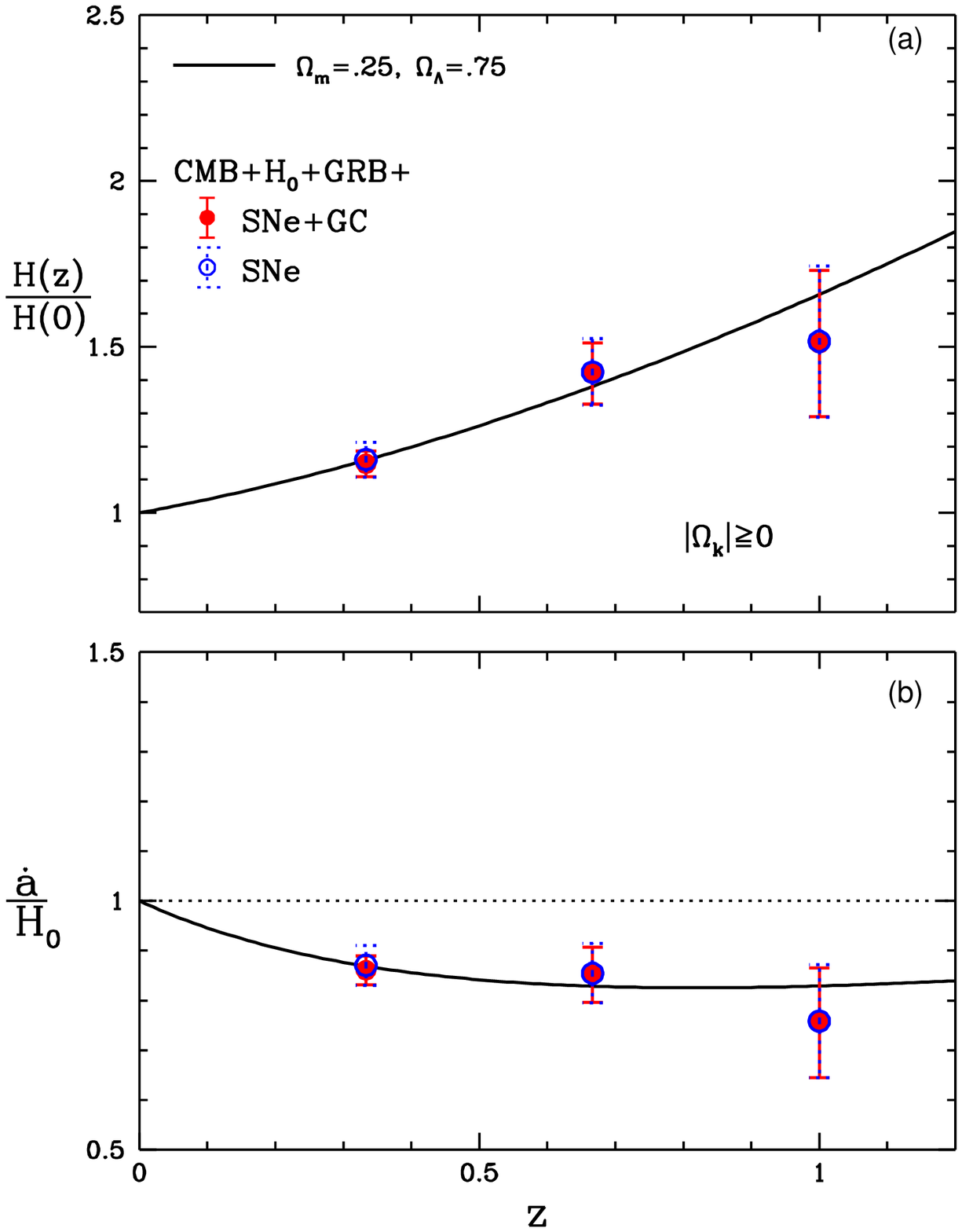,height=3.5in}
\hspace{-1in}
\psfig{file=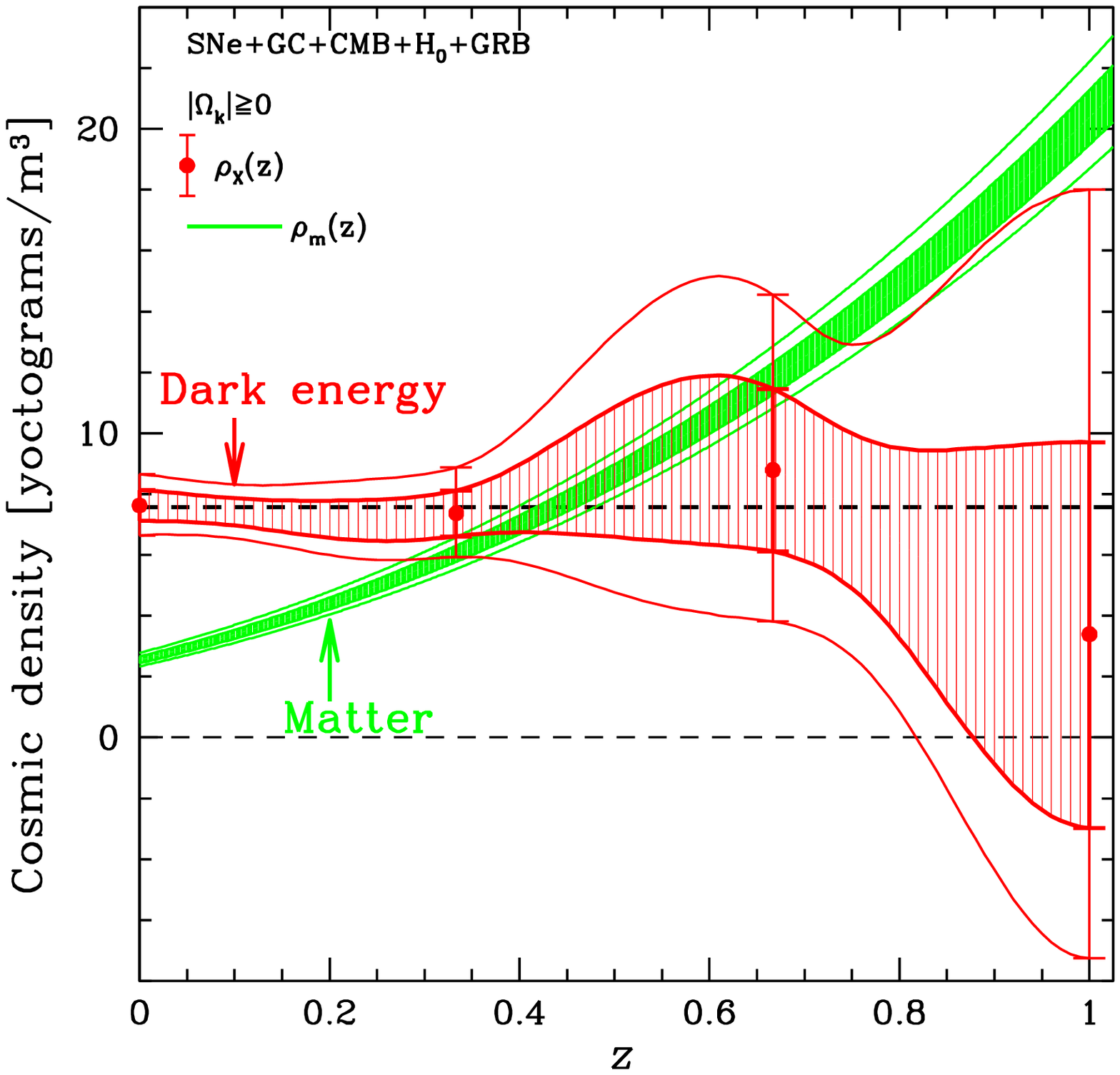,height=3.5in}
\vspace{-0.1in}
\caption{\footnotesize{Left panel: Expansion history of the universe
measured from current data \cite{Wang11}. Data used:
Cosmic microwave background anisotropy (CMB) data from
WMAP 7 year observations \cite{Komatsu11}; 
472 Type Ia supernovae (SNe Ia) (compiled by \cite{Conley11}, including 
data from the Supernova Legacy Survey (SNLS) \cite{Conley11},
the Hubble Space Telescope (HST) \cite{Riess07}, as well as nearby SNe Ia \cite{Hamuy96}); 
galaxy clustering measurements from Sloan Digital Sky Survey (SDSS) 
Luminous Red Galaxies (LRGs) \cite{Chuang11}, 69 Gamma Ray Bursts \cite{GRB},
and the latest Hubble constant ($H_0$) measurement using the Hubble Space Telescope (HST)
\cite{Riess11}.
Right panel: Dark energy density as a function of redshift measured from
the same data.
}}
\label{fig:Hz}
\end{figure}

Type Ia supernovae (SNe Ia), galaxy clustering (GC), and weak lensing (WL) are generally
considered the most powerful observational probes of dark energy.
SNe Ia provide a measurement of $H(z)$ derived from luminosity distances
of SNe Ia. GC provides direct measurements of $H(z)$, $D_A(z)$, as well
as the growth rate of cosmic large scale structure, $f_g(z)$.
WL provides measurements of $H(z)$ and the growth factor $G(z)$
(related to $f_g(z)$ via a derivative) \cite{wl}. It is important to utilize
all three methods, as they have different systematic
errors. Furthermore, GC and WL constrain different aspects in the
modification of gravity; both are required to achieve a robust
test of gravity.

Clusters of galaxies provide an independent and complementary method
to probe dark energy \cite{clusters}. Other methods, e.g., using gamma-ray bursts, 
old red galaxies, or radio galaxies \cite{de_other} provide additional 
cross-checks on dark energy constraints. CMB data and independent measurements of $H_0$ are
required to break the degeneracy between dark energy
and cosmological parameters (see e.g. \cite{WangPia07,Wang11}), 
hence are important as well in constraining dark energy.

Due to the page limit, I will focus on SNe Ia and GC, as these have 
yielded the strongest direct constraints on dark energy to date. 
I will first discuss the simplest general guidelines for probing dark energy, 
then SNe Ia and GC as dark energy probes respectively, 
and conclude with a brief summary of current status and future prospects of dark energy 
observational projects.

\section{Probing dark energy and testing gravity}

Because of the existence of two possible explanations, dark energy
and modified gravity, for the observed cosmic acceleration, it is
critical for us to recognize that we need to measure two functions
of cosmic time from observational data: the expansion history of the universe, 
$H(z)$, and the growth rate of cosmic large scale structure, $f_g(z)$.
Modified gravity models can give identical $H(z)$ as a dark energy
model by design, but the growth rate $f_g(z)$ is likely different
in these models compared to dark energy models.
The precise and accurate measurement of $H(z)$ and
$f_g(z)$ from observational data will allow us to probe the
true nature of cosmic acceleration \cite{Wang07}.

Dark energy is often parameterized by
a linear equation of state $w_X(a)=w_0+w_a (1-a)$ \cite{Chev01}.
Because of our ignorance of the nature of dark energy,
it is important to make model-independent constraints
by measuring the dark energy density $\rho_X(z)$
[or the expansion history $H(z)$] as a free function
of cosmic time.
Measuring $\rho_X(z)$ has advantages over measuring dark energy
equation of state $w_X(z)$ as a free function; $\rho_X(z)$ is more
closely related to observables, hence is more tightly 
constrained for the same number of redshift bins 
used \cite{WangGarna,rhox_other}.
Note that $\rho_X(z)$ is related to $w_X(z)$ as follows \cite{WangGarna}:
\begin{equation}
\label{eq:rhoprimew}
\frac{\rho_X(z)}{\rho_X(0)} = \exp\left\{ \int_0^z {\rm d}z'\, \frac{3
    [1+w_X(z')]}{1+z'} \right\}, 
\end{equation} 
{\it Hence parametrizing dark energy with $w_X(z)$ 
implicitly assumes that $\rho_X(z)$ does not
change sign in cosmic time.} 
This precludes whole classes of
dark energy models in which $\rho_X(z)$ becomes negative in the future
(``Big Crunch'' models, see \cite{Linde} for an example)\cite{WangTegmark04}.
If the present cosmic acceleration is caused by dark energy,
\be
E(z) \equiv \frac{H(z)}{H_0}=
\left[\Omega_m (1+z)^3 + \Omega_k (1+z)^2 +\Omega_X X(z)
\right]^{1/2},
\label{eq:E(z)}
\ee
where $X(z)\equiv \rho_X(z)/\rho_X(0)$. 
$H_0=H(z=0)$ is the Hubble constant.
$\Omega_m$ and $\Omega_X$ are the ratios of the matter and
dark energy density to the critical density $\rho_c^0=3H_0^2/(8\pi G)$,
and $\Omega_k=-k/H_0^2$ with $k$ denoting the curvature constant.
Consistency of Eq.(\ref{eq:E(z)}) at $z=0$ requires that 
$\Omega_m+\Omega_k+\Omega_X=1$.
Once $E(z)$ is specified, the evolution of matter density perturbations
on large scales, $\delta^{(1)}(\bfx,t)=D_1(t) \delta(\bfx)$ is determined
by solving the following equation for 
$D_1=\delta^{(1)}(\bfx,t)/\delta(\bfx)$,\footnote{Note that we 
have assumed that dark energy and dark matter are separate,
which is true for the vast majority of dark energy models
that have been studied in the literature. 
If dark energy and dark matter are coupled (a more complicated
possibility), or if dark energy and dark matter are unified
(unified dark matter models), Eq.(\ref{eq:fg}) would need 
to be modified accordingly.
Ref.\cite{Sandvik04} found the first strong evidence for the 
separation of dark energy and dark matter by ruling out a broad 
class of so-called unified dark matter models. They showed 
that these models produce oscillations or exponential blowup 
of the dark matter power spectrum inconsistent with observation.}
\be
\label{eq:fg}
D_1''(\tau) + 2E(z)D_1'(\tau) - {3\over 2}\Om (1+z)^3D_1 = 0,
\ee
where primes denote $d/d(H_0 t)$.
The linear growth rate 
\be
\label{eq:fgdef}
f_g(z) \equiv d\ln D_1/d\ln a.
\ee

In the simplest alternatives to dark energy, the present cosmic acceleration 
is caused by a modification to general relativity. 
Ref.\cite{modified_gravity_details} contains examples of
studies of observational signatures of modified gravity models.
A worked example is the DGP gravity model \cite{DGPmodel,DGP},
which can be described by a modified 
Friedmann equation\footnote{The validity of
the DGP model has been studied by \cite{DGP2}.}.
The right panel of Fig.{\ref{fig:Hzfg}} shows a DGP model
that gives identical $H(z)$ as a dark energy model, but
gives significantly different $f_g(z)$ \cite{Wang07}.

\section{Type Ia supernovae as dark energy probe}

The use of Type Ia supernovae (SNe Ia) is the best established method for 
probing dark energy, since this is the method through which cosmic acceleration
has been discovered \cite{de}. This method is independent of the clustering of
matter\footnote{Galaxy peculiar velocities from large-scale supernova
surveys can be used to probe dark energy \cite{Bhattacharya11}.}, and can provide a robust measurement of 
$H(z)$ \cite{WangTegmark05}
through the measured luminosity distance as a function of redshift,
$d_L(z)=(1+z)\, r(z)$, where the comoving distance $r(z)$ 
from the observer to redshift $z$ is given by
\ba
\label{eq:rz}
&&r(z)=cH_0^{-1}\, |\Omega_k|^{-1/2} {\rm sinn}[|\Omega_k|^{1/2}\, \Gamma(z)],\\
&&\Gamma(z)=\int_0^z\frac{dz'}{E(z')}, \hskip 1cm E(z)=H(z)/H_0 \nonumber
\ea
where ${\rm sinn}(x)=\sin(x)$, $x$, $\sinh(x)$ for 
$\Omega_k<0$, $\Omega_k=0$, and $\Omega_k>0$ respectively.\\

\begin{figure} 
\psfig{file=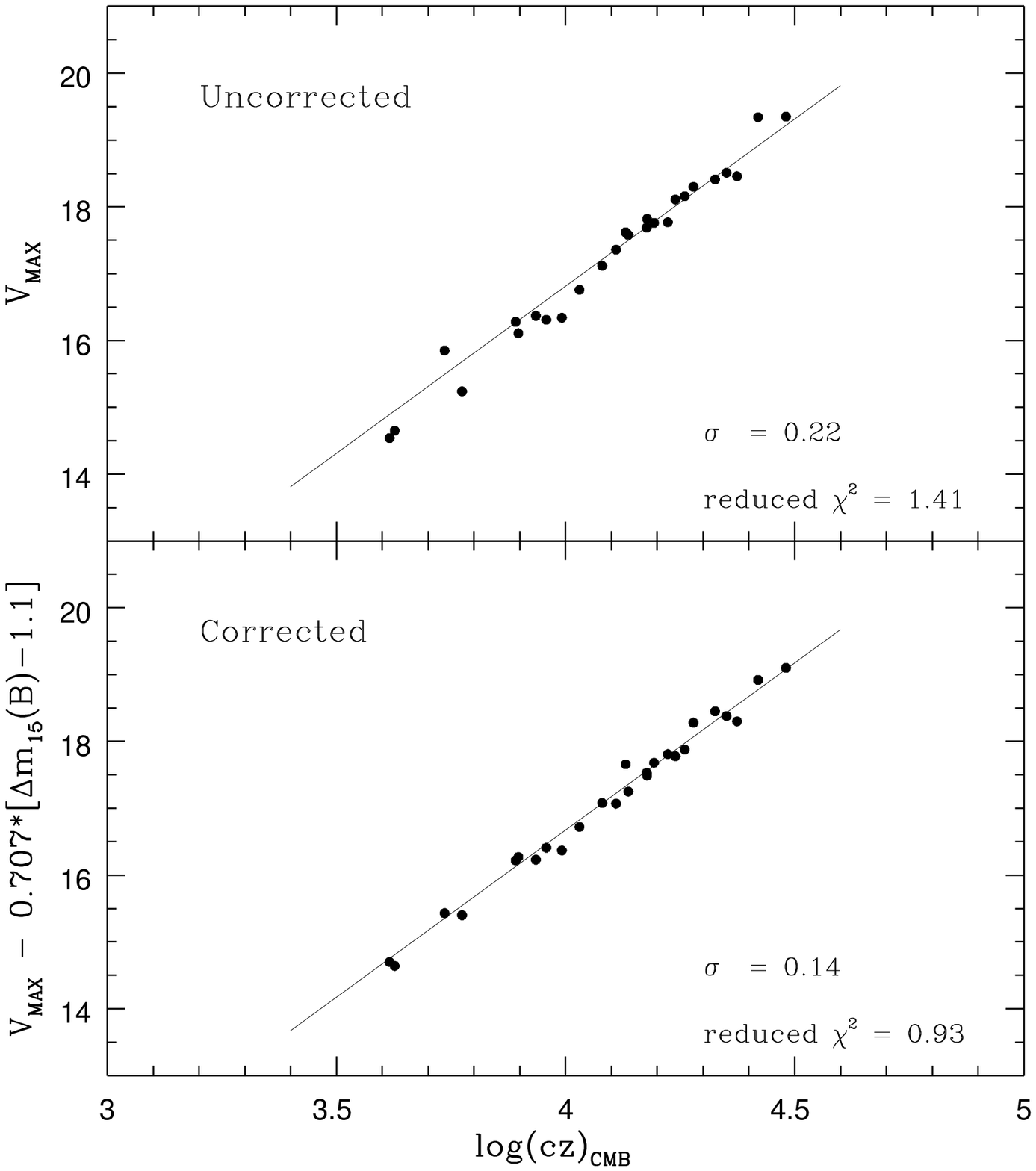,height=3.5in}
\hspace{-0.1in}
\psfig{file=jhk_hubble.ps,height=2.8in}
\vspace{-0.1in}
\caption{\footnotesize{Left panel: Hubble diagrams showing 26 SNe Ia
with $B_{max}-V_{max} \leq 0.20$ from the Calan/Tololo sample \cite{Hamuy96}.
This sample provided half of the data for the discovery of the 
cosmic acceleration in 1998 \cite{de}. 
The solid lines indicate Hubble's law;
perfect standard candles (with $\sigma=0$) fall on these lines.
Right panel: Hubble diagrams of SNe Ia in the NIR bands.
Note that these SNe Ia have only been corrected for
dust extinction; {\it no} corrections have been made
for lightcurve width. \cite{KK04}
}}
\label{fig:sn_cali}
\end{figure}

{\bf Type Ia as standard candles.}
A SN~Ia is a thermonuclear explosion that
completely destroys a carbon/oxygen white dwarf near
the Chandrasekher limit of 1.4 $M_{\odot}$.
This is the reason SNe~Ia are so uniform in peak luminosity.
The first challenge to overcome when using SNe~Ia as cosmological
standard candles is properly incorporating the intrinsic scatter in SN~Ia peak luminosity.
The usual calibration of SNe~Ia reduces the intrinsic scatter in SN~Ia peak 
luminosity (Hubble diagram dispersion) to about 0.16~mag 
\cite{Phillips93,Riess95}.
The calibration techniques used so far are based on one observable parameter,
the lightcurve width, which can be parametrized either as $\Delta m_{15}$
(decline in magnitudes for a SN~Ia in the first 15 days after $B$-band maximum,
see \cite{Phillips93}), or a stretch factor (which linearly scales the time axis, 
see \cite{Goldhaber01}).
The lightcurve width is associated with the amount of $^{56}$Ni
produced in the SN~Ia explosion, which in turn depends on
when the carbon burning makes the transition from turbulent 
deflagration to a supersonic detonation \cite{Wheeler03}.
There may be additional physical parameters associated with
SN Ia lightcurves or spectra \cite{sn_other}
that can further improve the calibration of SNe Ia.
Fig.\ref{fig:sn_cali} (left panel) shows a historic example of
the homogeneity of SNe Ia \cite{Hamuy96}.\\

{\bf Systematic effects of SNe Ia as dark energy probe.}
The main systematic effects for using SNe Ia to probe dark energy
are: extinction by normal \cite{Card89}
or gray dust \cite{Aguirre99}\footnote{Gray dust, consisting of large dust grains,
is difficult to detect by its reddening and could mimic
the effect of dark energy \cite{Aguirre99}. 
Gray dust can be constrained quantitatively by the Cosmic Far 
Infrared Background \cite{Aguirre00},
with no evidence found in favor of gray dust so far.
Supernova flux correlation measurements can be used in 
combination with other lensing data to infer the level of dust 
extinction, and provide a viable method to 
eliminate possible gray dust contamination
in SN Ia data \cite{Zhang07}.}, weak lensing amplification by cosmic
large scale structure \cite{sn_weak lensing}, and possible evolution in the 
peak luminosity of SNe Ia.

Recent data show that the apparent dust extinction of
SNe Ia is very different from the typical extinction law
due to Milky Way dust, possibly due to
the mixing of intrinsic SN Ia color variation with
dust extinction, or variations
in the properties of dust \citep{Conley07}.
The extinction by dust can be corrected using
multi-band imaging data, especially near infrared (NIR) observations
of SNe, since dust extinction decreases with wavelength.

\begin{figure} 
\psfig{file=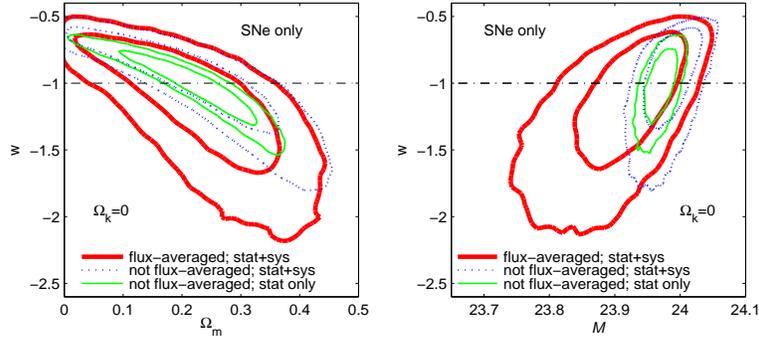,width=4in}\\
\caption{\label{fig:sn_flux}\footnotesize%
The 2D marginalized contours of $(w, \Omega_m, {\cal M})$
for SNe data compiled by \cite{Conley11} (with and without flux-averaging), assuming a flat universe.
The contours are at 68\% and 95\% confidence levels. \cite{Wang11}
}
\end{figure}

\begin{figure} 
\psfig{file=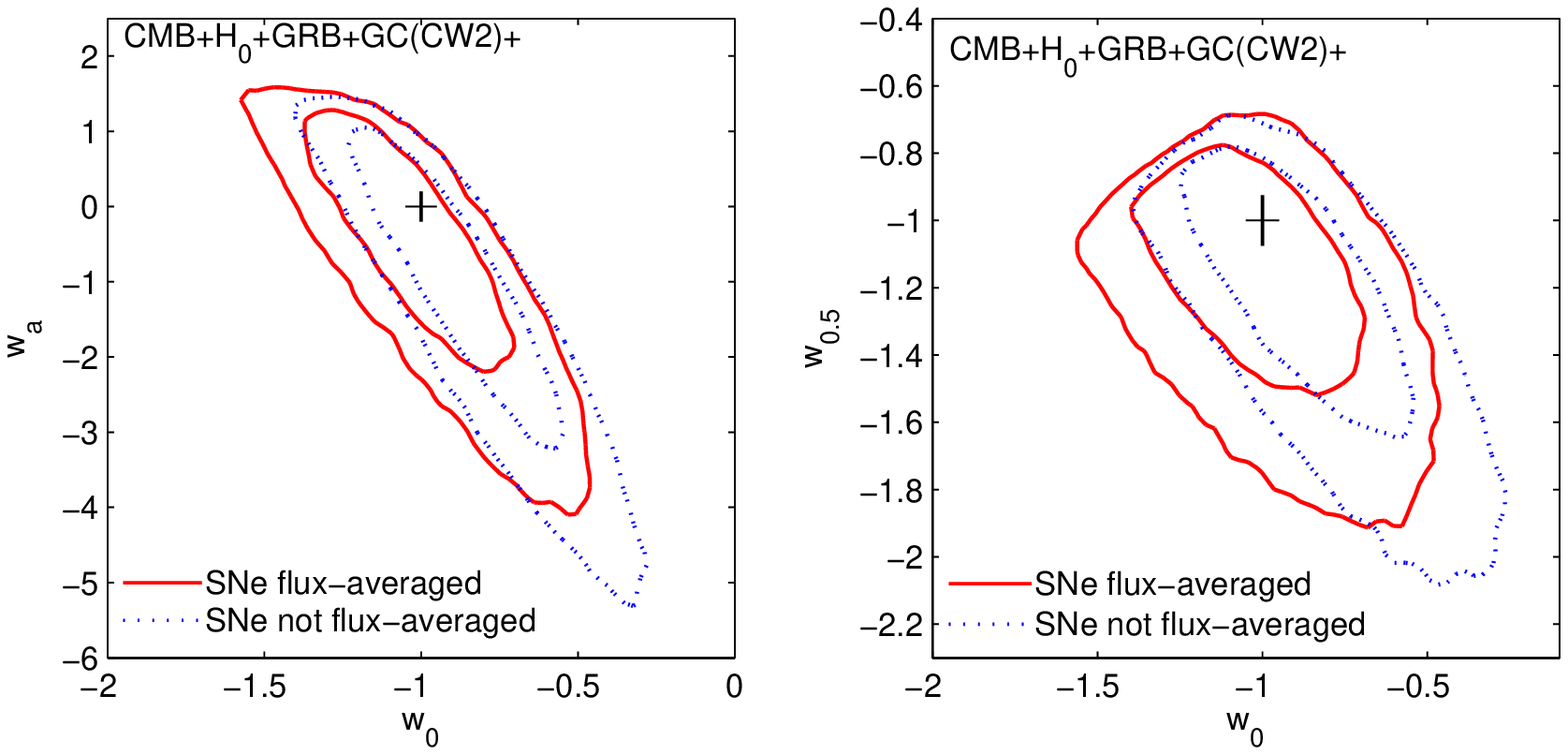,width=4in}\\
\caption{\label{fig:w0w1_flux_GC}\footnotesize%
The 2D marginalized contours of $(w_0,w_a)$ and $(w_0,w_{0.5})$
for SNe data (with and without flux-averaging) combined with galaxy 
clustering (CW2), CMB, $H_0$, and GRB data (same data as in Fig.\ref{fig:Hz}).
The contours are at 68\% and 95\% confidence levels. \cite{Wang11}
}
\end{figure}

\begin{figure} 
\psfig{file=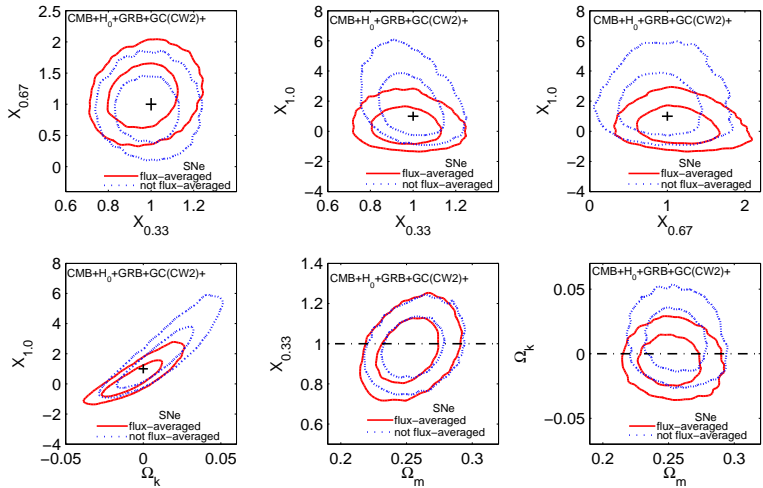,width=4in}\\
\caption{\label{fig:X3_flux_GC}\footnotesize%
The 2D marginalized contours of $(X_{0.33}, X_{0.67}, X_{1.0}, \Omega_m, \Omega_k)$
for SNe data (with and without flux-averaging) combined with galaxy clustering (CW2), 
CMB, $H_0$, and GRB data (same data as in Fig.\ref{fig:Hz}).
The contours are at 68\% and 95\% confidence levels. \cite{Wang11}
}
\end{figure}

The weak lensing amplification of SNe Ia by cosmic
large scale structure can be modeled by a universal 
probability distribution function for 
weak-lensing amplification based on the measured
matter power spectrum \cite{Wang02}.
The effect of weak lensing on the SN Ia data
can be minimized through flux-averaging \cite{flux-averaging}.
Figs.\ref{fig:sn_flux} shows the 2D marginalized contours of
$(w, \Omega_m, {\cal M})$ (where ${\cal M})$ is a nuisance parameter), 
assuming a constant equation of state for dark energy, $w$, and a flat universe.
Note that the inclusion of systematic errors of SNe 
leads to significantly larger uncertainties in
estimated parameters, compared to
when only statistical errors of SNe are included \cite{Wang11}.
Clearly, flux-averaging (thick solid lines) leads to 
larger errors on dark energy and 
cosmological parameters if only SN Ia data are used. 
However, when other data are added, flux-averaging leads
to smaller errors on dark energy (see Figs.\ref{fig:w0w1_flux_GC}-\ref{fig:X3_flux_GC})
because flux-averaging increases the concordance of
SNe Ia with other data.

The evolution in SN Ia peak luminosity could arise
due to progenitor population drift, since the most distant
SNe Ia come from a stellar environment very different 
(a much younger universe) than that of the nearby SNe Ia.
However, with sufficient statistics, we can subtype SNe Ia
and compare SNe Ia at high redshift and low redshift 
that are similar in both lightcurves and spectra,
thus overcoming the possible systematic effect due to
progenitor population drift \cite{Branch01}.\\

\begin{figure} 
\psfig{file=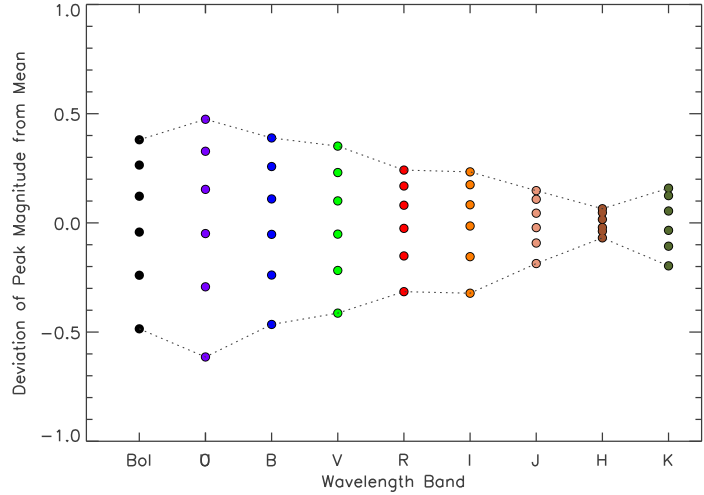,height=2.8in}
\caption{\footnotesize{Dispersion in peak magnitude (measured at the first
lightcurve maximum) as a function of wavelength band for SN Ia models
with \Nifs\ masses between 0.4 and 0.9~\msun (Kasen 2006 \cite{Kasen06}).
}}
\label{fig:Kasen_f15}
\end{figure}

{\bf Optimized observations of SNe Ia.}
NIR observations of SNe Ia provide additional strong advantages beyond
being relatively dust-free.
SNe Ia are better standard candles at NIR wavelengths 
compared to the optical wavelengths \cite{KK04,Phillips06,WV07}.
The right panel of Fig.{\ref{fig:sn_cali}} shows the Hubble diagram of SNe Ia in 
the NIR, {\it without} the usual lightcurve width correction. 
The smaller intrinsic dispersion of SN Ia peak luminosity
in the NIR can be explained by the theoretical modeling of
SN Ia lightcurves using time-dependent multi-group
radiative transfer calculations (Kasen 2006 \cite{Kasen06}).
Fig.{\ref{fig:Kasen_f15}} shows the dispersion in peak magnitude 
(measured at the first lightcurve maximum) as a function of wavelength 
band for SN Ia models with \Nifs\ masses between 0.4 and 0.9~\msun
\cite{Kasen06}.

It is important to obtain high quality spectra of SNe Ia (including
NIR spectra, see \cite{Marion06_09}), since the spectra of SNe Ia have been 
shown to provide calibration relations that decrease the scatter of SNe Ia 
in the Hubble diagram, and make SNe Ia better distance indicators.
The correlation between SN Ia spectroscopic features and
luminosity has been found in the observational data (see, e.g.,
\cite{Garnavich04}). More recently, Bailey et al. (2009) \cite{Bailey09} used the Nearby 
Supernova Factory spectrophotomery of 58 SNe Ia to perform 
an unbiased search for flux ratios that correlate with SN Ia 
luminosity. They found that the 642/443 nm flux ratio 
is most strongly correlated with SN Ia absolute magnitudes.
The correlation of SN Ia spectroscopic features and
luminosity can be understood through comparing theoretical
modeling with observational data.
Hachinger et al. (2008) \cite{Hachinger08} found that the strength 
of the Si II $\lambda$5972 line may be a very promising spectroscopic 
luminosity indicator for SNe Ia, with the correlation between Si II $\lambda$5972 
strength and luminosity resulting from the effect of ionization balance.

\begin{figure} 
\psfig{file=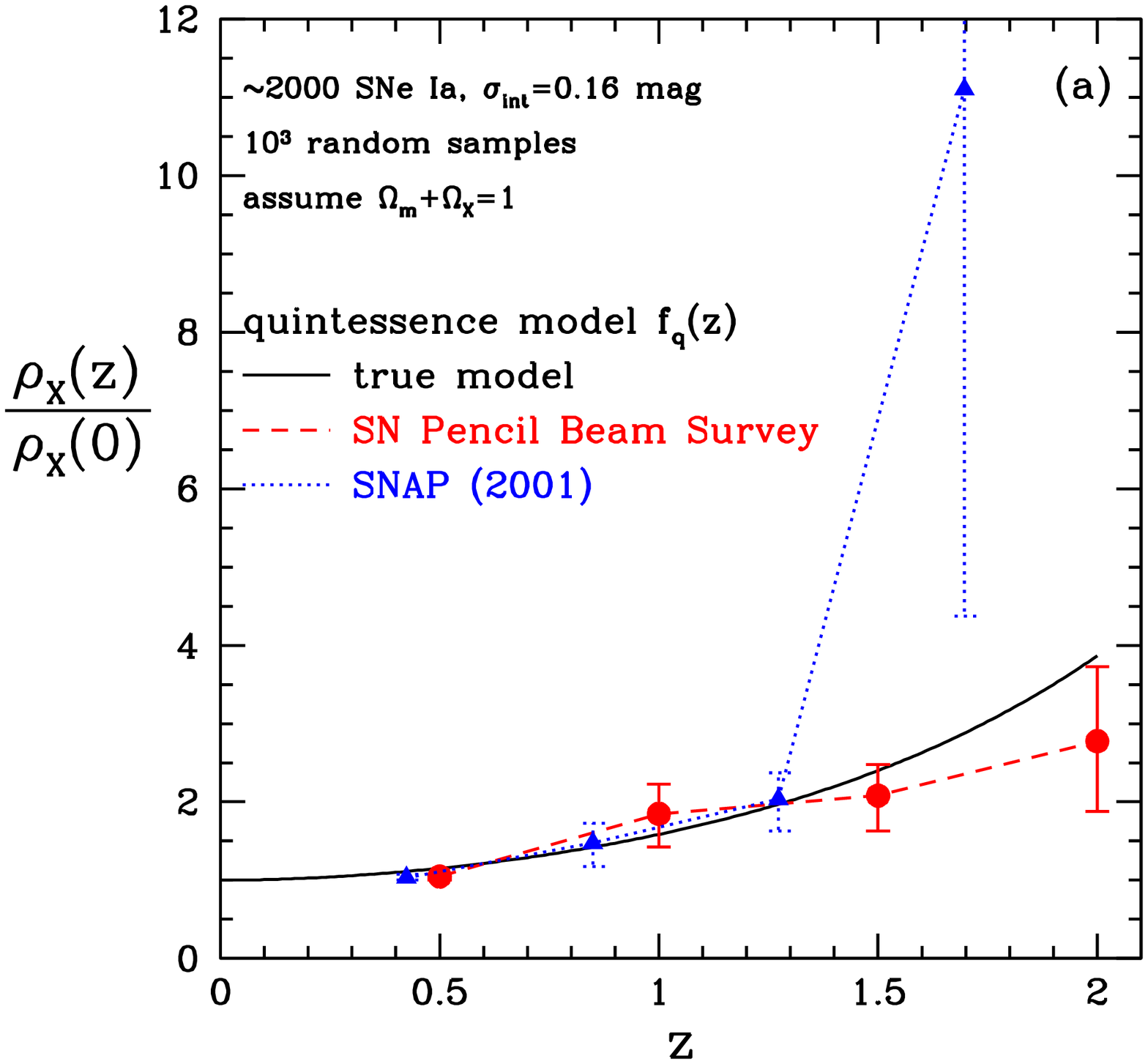,height=2.6in}
\psfig{file=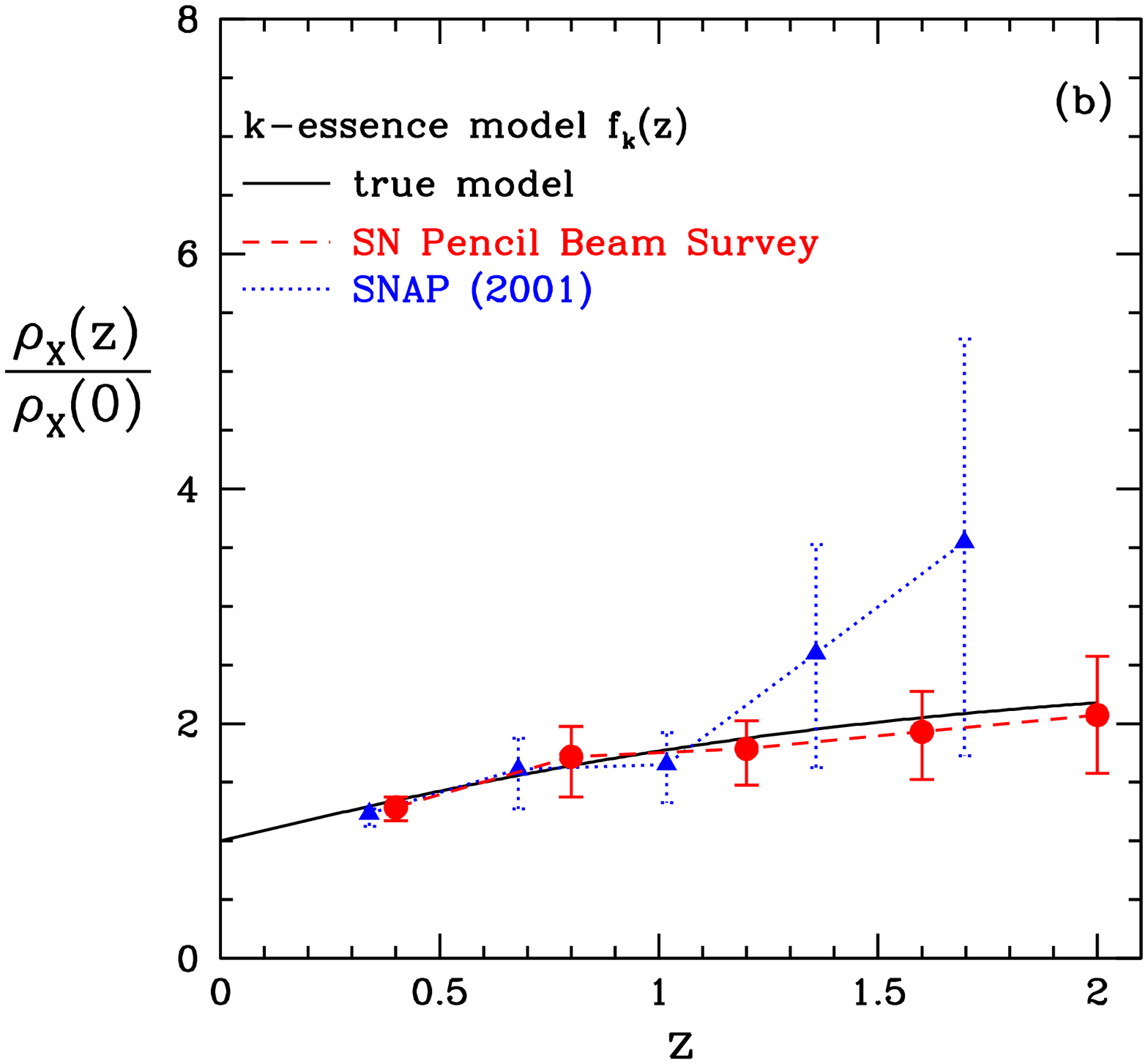,height=2.6in}
\vspace{-1.5in}
\caption{\footnotesize{The comparison of an ultra deep supernova survey
\cite{Wang2000a} with a much shallower survey
in the reconstruction of the dark energy density $\rho_X(z)$ as
a free function of cosmic time \cite{WangLovelace2001}.
}}
\label{fig:sn_deep}
\end{figure}

The key to the efficient use of SNe Ia for probing dark energy is
to obtain the largest possible unbiased sample of SNe Ia at the greatest
distances from the observer \cite{WangLovelace2001}.
This is achieved by an ultra deep survey of the same areas in
the sky every few days over at least one year \cite{Wang2000a}.
Given the same observational resources, an ultra deep supernova survey
is superior to a much shallower survey. A sufficiently deep supernova survey 
is required to reconstruct the dark energy density $\rho_X(z)$ as
a free function of cosmic time (i.e., to measure $H(z)$ precisely,
see Fig.\ref{fig:sn_deep}) \cite{WangLovelace2001}.

\section{Galaxy clustering as dark energy probe}

\begin{figure} 
\psfig{file=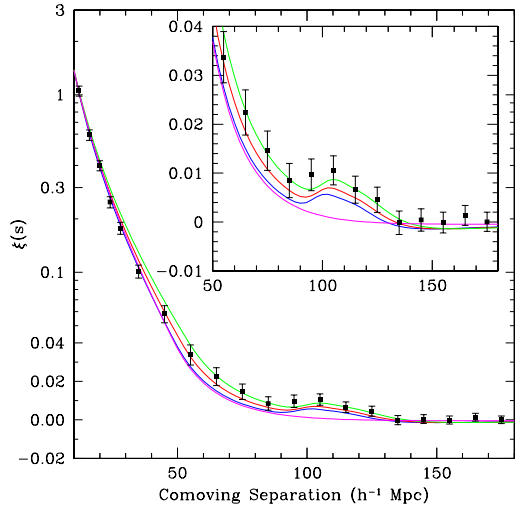,height=2.8in}
\caption{\footnotesize{The spherically-averaged galaxy correlation function measured 
from the SDSS data, clearly showing a peak corresponding to the
BAO scale at $\sim$ 100$\,h^{-1}$Mpc \cite{Eisen05}.
}}
\label{fig:BAO}
\end{figure}

{\bf Baryon acoustic oscillations (BAO) as a standard ruler.}
At the last scattering of CMB photons, the acoustic oscillations 
in the photon-baryon fluid became frozen, and imprinted their signatures
on both the CMB (the acoustic peaks in the CMB angular
power spectrum) and the matter distribution (the baryon acoustic
oscillations in the galaxy power spectrum).
Because baryons comprise only a small fraction of matter,
and the matter power spectrum has evolved significantly
since last scattering of photons, BAO are much smaller in amplitude
than the CMB acoustic peaks, and are washed out on small
scales. BAO in the observed galaxy power spectrum have the characteristic 
scale determined by the comoving sound horizon at the drag epoch
(which occurred shortly after recombination), which is precisely 
measured by the CMB anisotropy data (see, e.g., \cite{Komatsu11}).
Comparing the observed BAO scales with the expected values
gives $H(z)$ in the radial direction, and $D_A(z)=r(z)/(1+z)$ (the angular
diameter distance) in the transverse direction \cite{bao_first,BAO}.
Fig.{\ref{fig:BAO}} shows the first detection of the BAO peak
from a sample of the SDSS LRGs \cite{Eisen05}.

BAO represents only a fraction of the cosmological information
contained in galaxy clustering data. A flux-limited galaxy redshift survey can
allow us to measure the cosmic expansion history $H(z)$ through BAO only or the 
shape of the galaxy power spectrum $P(k)$ (which includes BAO as features)
measured from the galaxy distribution, and the growth history of
cosmic large scale structure $f_g(z)$ through independent measurements of 
redshift-space distortions and the bias factor between the distribution of galaxies
and that of matter \cite{Wang07}. Assuming linear bias, the combination
$f_g(z)\sigma_{8m}(z)$ can be measured directly and used to test gravity \cite{Song09}.\\

\begin{figure} 
\psfig{file=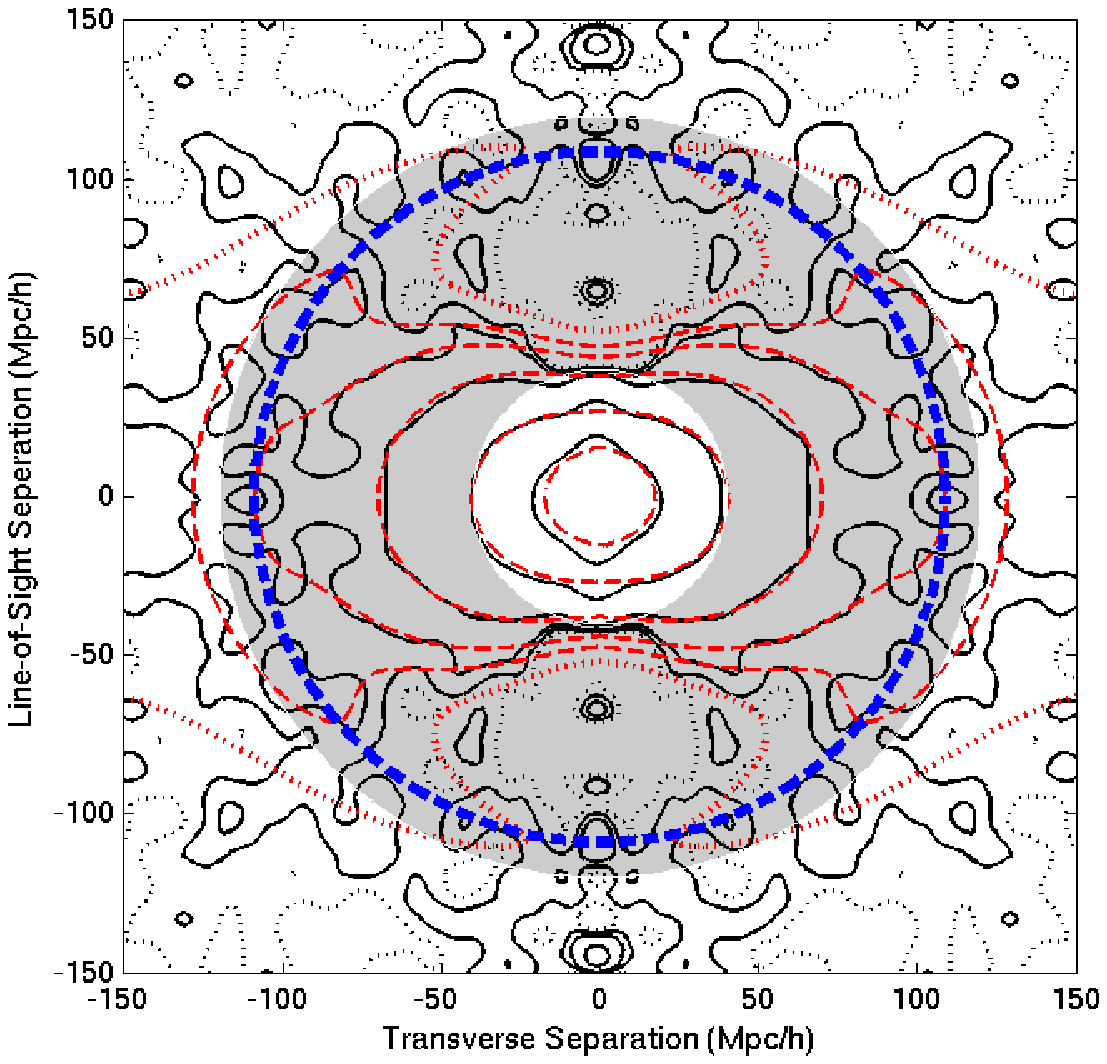,height=2.8in}
\hspace{-1in}
\psfig{file=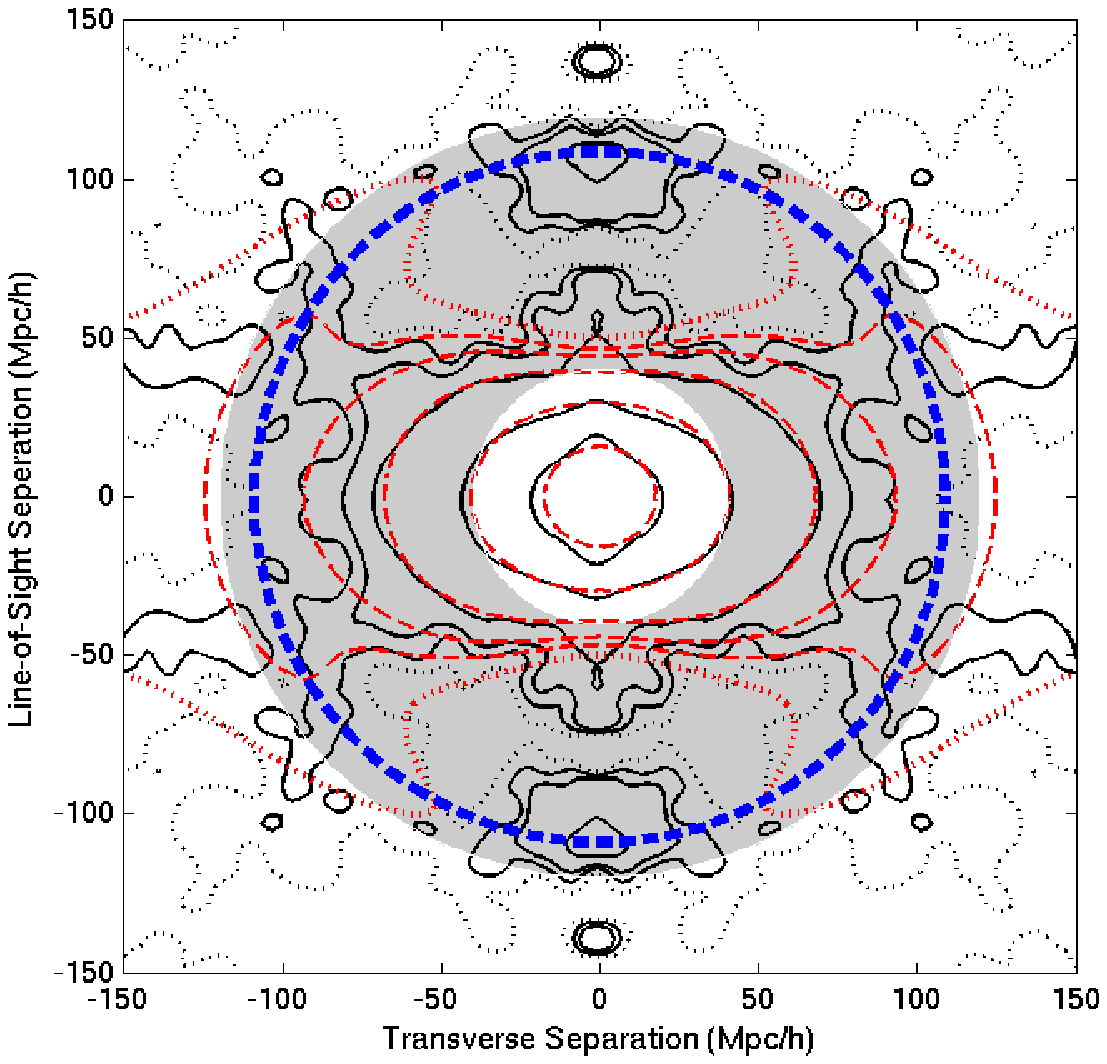,height=2.8in}
\caption{\footnotesize{The two-dimensional two-point correlation function (2D 2PCF)
measured from SDSS DR7 LRGs (left panel) and a LasDamas SDSS LRG mock catalog (right panel)
in a redshift range  $0.16 < z < 0.44$ (solid black
contours), compared to a theoretical correlation function with parameters close to the 
best fit values in the likelihood analysis (dashed red contours). 
In both figures, the shaded disk indicates the scale range considered ($s=40-120\ h^{-1}$Mpc ) in this study.
The thick dashed blue circle denotes the baryon acoustic oscillation scale. 
The observed 2D 2PCF has been smoothed by a Gaussian filter with rms variance 
of $2 h^{-1}$Mpc for illustration in these figures only; smoothing is not used in our likelihood 
analysis. The contour levels are $\xi=0.5, 0.1, 0.025, 0.01, 0.005, 0$. 
The $\xi=0$ contours are denoted with dotted lines for clarity. \cite{Chuang11}
}}
\label{fig:sdss_2d}
\end{figure}

{\bf Current GC Measurements.}
The first simultaneous measurements of $H(z)$ and $D_A(z)$ from galaxy clustering
data was made very recently by Chuang \& Wang (2011) \cite{Chuang11}, based
on a Markov Chain Monte Carlo (MCMC) analysis of the two-dimensional two-point correlation 
function (2D 2PCF) they measured from the flux-limited sample of LRGs from the SDSS Data 
Release 7 \cite{Chuang11}.
Fig.\ref{fig:sdss_2d} shows the 2D 2PCF measured from the SDSS LRGs and a single 
LasDamas SDSS LRG mock catalog for comparison. The similarity between the
data and the mock in the range of scales used (indicated by the shaded
disk) is apparent. Due to the current limitations in the modeling of
systematic effects, only the quasi-linear scale range of $s=40-120\,h^{-1}$Mpc is used
for a conservative estimate in this analysis.
Chuang \& Wang (2011) \cite{Chuang11} obtained
\be
H(z=0.35)=82.1_{-4.9}^{+4.8}\, \Hunit, \hspace{0.2in}
D_A(z=0.35)=1048_{-58}^{+60}\, \mbox{Mpc}
\ee
without assuming a dark energy model or a flat universe. Scaling $H(z)$ and $D_A(z)$
using $r_s(z_d)$ (the sound horizon at the drag epoch) in the MCMC analysis, they found
that the derived measurements of 
\be
H(0.35)r_s(z_d)=13020\pm 530 \,\mbox{km/s}, \hspace{0.2in}
r_s(z_d)/D_A(0.35)=0.1518\pm 0.0062
\label{eq:HD_A}
\ee
are nearly uncorrelated (with a normalized correlation coefficient of $r = -0.0584$), 
have tighter constraints and are more robust with respect to possible systematic effects. 
This is as expected, since $H(0.35) \,r_s(z_d)$ and $r_s(z_d)/D_A(0.35)$
correspond to the preferential redshift separation along the line of sight,
and the preferential angular separation in the transverse direction respectively;
these in turn arise from the BAO in the radial and transverse directions.
The measurable preferential redshift and angular separations
should be uncorrelated since they are independent degrees of freedom.
The presence of the BAO (although only marginally visible in Fig.\ref{fig:sdss_2d})
leads to tight and robust constraints on $H(0.35) \,r_s(z_d)$ and $r_s(z_d)/D_A(0.35)$.
Since most of the constraining power in this analysis comes from fitting 
the overall shape of the galaxy correlation function on quasi-linear scales, 
and not from fitting the BAO peaks, these measurements are galaxy
clustering measurements (rather than BAO only measurements).

The constraints in Eq.(\ref{eq:HD_A}) can be used to combine with cosmic microwave background and any other 
cosmological data sets to constrain dark energy, as {\it no} priors were imposed that would affect the
combined constraints \cite{Chuang10,Chuang11}. These results have significant implications for future 
surveys in establishing the feasibility of measuring both $H(z)$ and $D_A(z)$ from galaxy clustering data. 

\begin{figure} 
\psfig{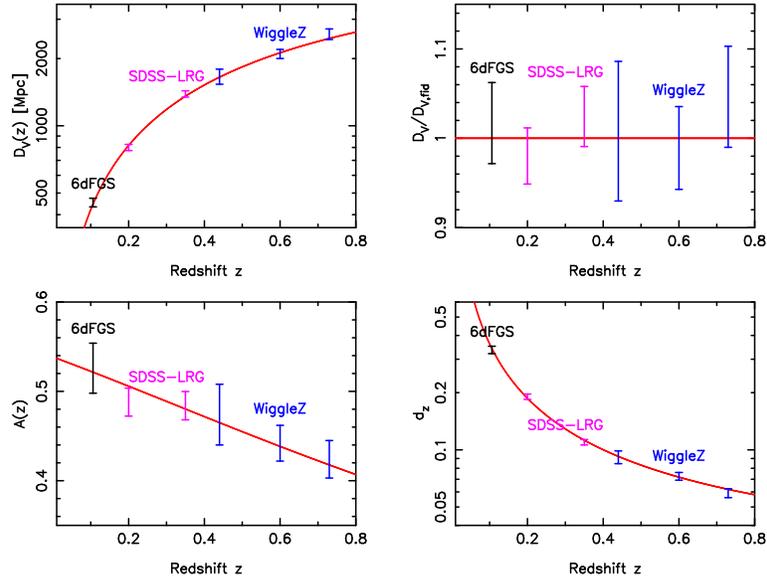}
\caption{\footnotesize{Measurements of the baryon acoustic peak at redshifts z= 0.44, 0.6 and 0.73 in the galaxy 
correlation function of the final data set of the WiggleZ Dark Energy Survey. \cite{Blake11a}
}}
\label{fig:Blake_bao}
\end{figure}

\begin{figure} 
\psfig{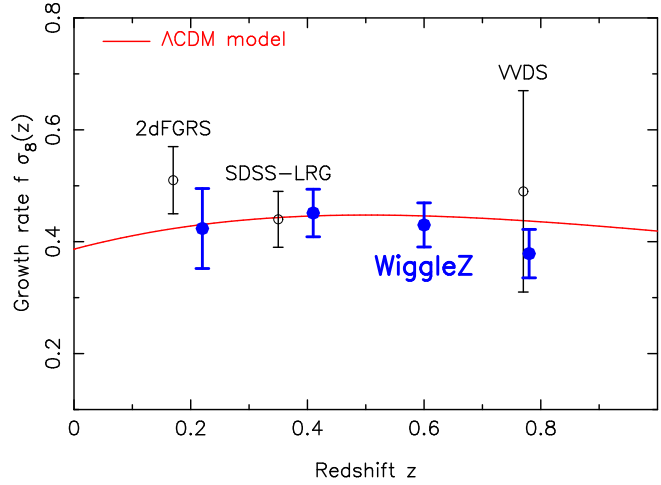}
\caption{\footnotesize{Measurements of $f_g(z)\sigma_{8m}(z)$ from the WiggleZ Dark Energy Survey data. \cite{Blake11b}
}}
\label{fig:Blake_fgs8}
\end{figure}

Fig.\ref{fig:Blake_bao} and Fig.\ref{fig:Blake_fgs8} shows the first results from the WiggleZ Dark Energy 
Survey obtained by Blake et al. (2011ab) \cite{Blake11a,Blake11b}. Note that Fig.\ref{fig:Blake_bao} 
assumes fixed values of ($\Omega_b h^2$, $n_S$, $h$, $\sigma_8$), and the background cosmological
model is assumed to be known for Fig.\ref{fig:Blake_fgs8}.
When dramatically larger data sets become available from the next generation galaxy redshift surveys, 
it will be possible to extract both distance and growth rate
measurements simultaneously without imposing strong priors (see, e.g., \cite{Euclid_RB}).
\\

\begin{figure} 
\psfig{file=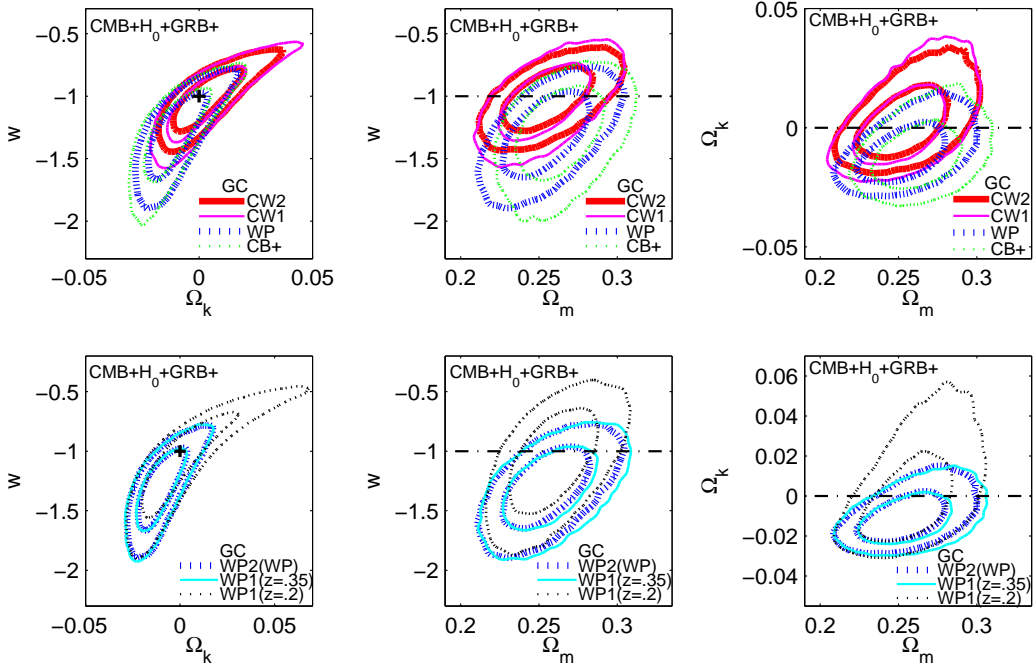,width=5.5in}\\
\caption{\label{fig:w_GC}\footnotesize%
The 2D marginalized contours of $(w, \Omega_m, \Omega_k)$
for different galaxy clustering measurements combined with 
CMB, $H_0$, and GRB data.
The contours are at 68\% and 95\% confidence levels.
}
\end{figure}

{\bf Possible systematic differences in different GC measurements.}
Ref.\cite{Wang11} showed that there may be systematic differences in different GC measurements.
Fig.\ref{fig:w_GC} shows the 2D marginalized contours of $(w, \Omega_m, \Omega_k)$
for different GC measurements combined with 
CMB, $H_0$, and GRB data.
The first row of Fig.\ref{fig:w_GC} compares the
$H(z=0.35)r_s(z_d)$ and $r_s(z_d)/D_A(z=0.35)$ measurements by 
Chuang \& Wang (2011) \cite{Chuang11} with their $d_{0.35}=r_s(z_d)/D_V(z=0.35)$ 
measurement (both from SDSS DR7 LRGs), as well as the
$d_{0.2}$ and $d_{0.35}$ measurements by Percival et al. (2010) 
\cite{Percival10} from SDSS DR7 LRG and main galaxy samples and 2dFGRS,
and the $d_{0.6}$ measurement by Blake et al. (2011) from
the WiggleZ survey \cite{Blake11} combined with the
$d_{0.106}$ measurement by Beutler et al. (2011) from 6dF GRS \cite{6dF}.

For the Chuang \& Wang (2011) \cite{Chuang11} GC measurements (CW2
and CW1), the constraints on $w$ are tightened significantly by going 
from spherically-averaged data (CW1), i.e., $d_{0.35}$, to 2D data (CW2), i.e.,
$H(z=0.35)r_s(z_d)$ and $r_s(z_d)/D_A(z=0.35)$, as indicated
by comparing the thin solid contours (CW1) to thick solid contours 
(CW2) in the first row of Fig.\ref{fig:w_GC}.
This is as expected, as more information from GC is included in CW2
compared to CW1.
Both the Percival et al. (2010) GC measurements (WP) and the
combined WiggleZ survey and 6dF GRS measurements (CB+)
favor $w<-1$ (similar results were found by \cite{GC_other} 
using GC measurements from \cite{Percival10}), while the Chuang \& Wang (2011) 
\cite{Chuang11} GC measurements favor $w=-1$.

The second row in Fig.\ref{fig:w_GC} compares the 
$d_{0.2}$ and $d_{0.35}$ measurements by Percival et al. (2010) 
\cite{Percival10} (WP2), with their measurements of
$d_{0.2}$ and $d_{0.35}$ separately.
Clearly, most of the constraining power on $w$ comes from 
$d_{0.35}$. While the $d_{0.2}$ measurement favors
$w=-1$, the $d_{0.35}$ measurement favors $w<-1$.

The measurements of $d_{0.35}$ by Chuang \& Wang (2011) \cite{Chuang11}
and Percival et al. (2010) \cite{Percival10} are similar
in precision, but differ systematically:
$d_{0.35}^{CW} \equiv r_s(z_d)/D_V(z=0.35)=0.1161\pm   0.0034$,
while $d_{0.35}^{WP}\equiv r_s(z_d)/D_V(z=0.35)=0.1097\pm 0.0036$.
The lower measured value of $d_{0.35}^{WP}$ implies 
a smaller $H(z=0.35)$, which in turn implies a more
negative $w$. When combined with CMB, $H_0$, and GRB data,
$d_{0.35}^{CW}$ favors $w=-1$, while $d_{0.35}^{WP}$ favors 
$w<-1$. Note that these two measurements used different
methods to analyze GC data: Chuang \& Wang (2011) used
the galaxy correlation function, while Percival et al. (2010)
used galaxy power spectrum. It is not surprising that
they lead to different distance measurements from GC.\\

\begin{figure}
\psfig{file=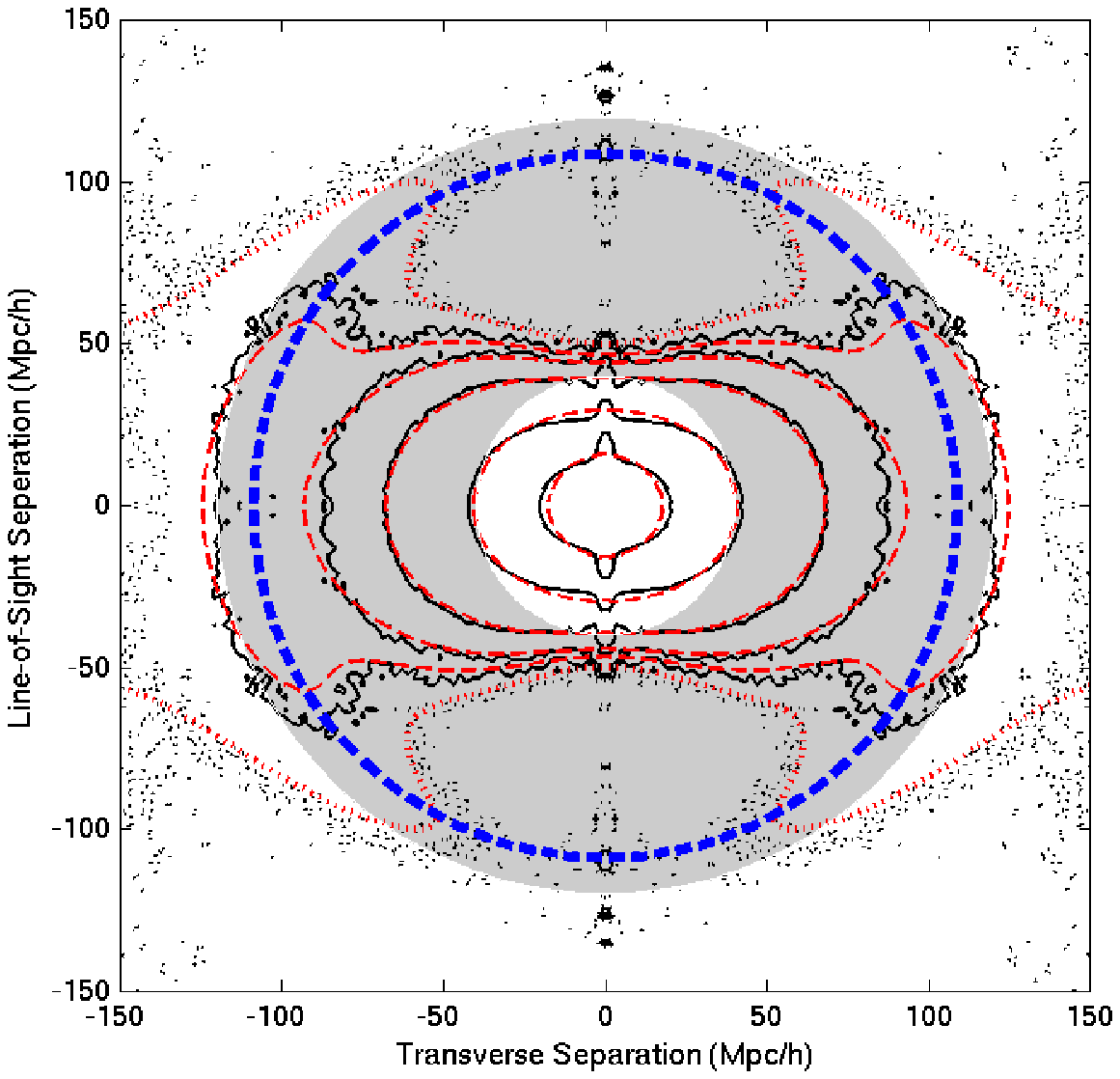,height=2.8in}
\hspace{-0.5in}
\psfig{file=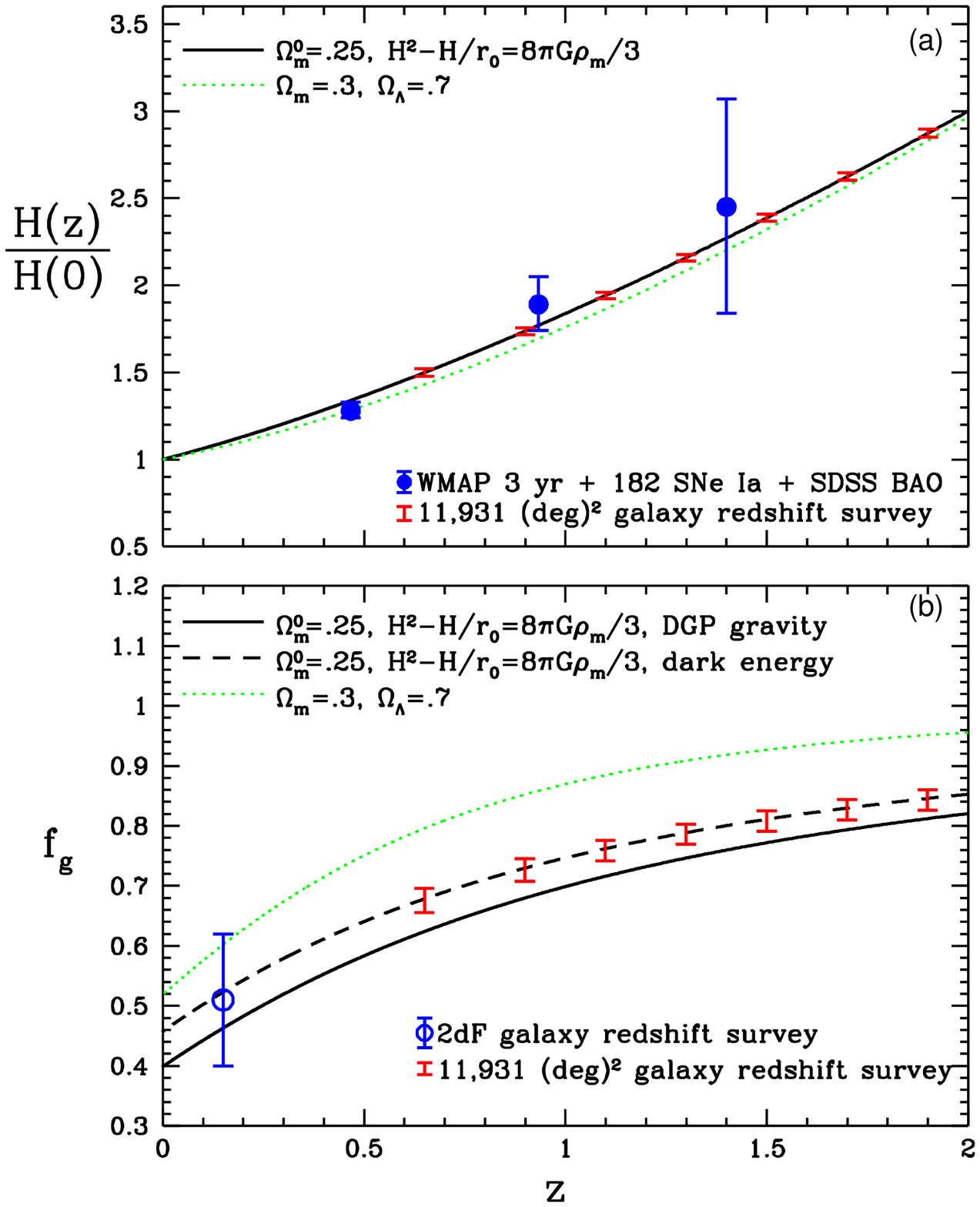,height=3.2in}
\caption{\footnotesize{Left panel: The average two-dimensional two-point correlation
function (2D 2PCF) measured from 160 LasDamas SDSS LRGfull mock catalogs (solid black
contours), compared to a theoretical model with the input parameters 
of the LasDamas simulations (dashed red contours). The line types and
contour levels are the same as in Fig.\ref{fig:sdss_2d} \cite{Chuang11}.
Right panel: Current and expected future measurements of the 
cosmic expansion history $H(z)=H_0 E(z)$ and the growth rate of 
cosmic large scale structure $f_g(z)$). The future data correspond to 
a magnitude-limited NIR galaxy redshift survey
covering $>$10,000 square degrees and $0.5<z<2$. \cite{Wang07} }}
\label{fig:Hzfg}
\end{figure}

{\bf Future Prospects.}
The data contours in the left panel of Fig.\ref{fig:Hzfg} gives a sense of the
exquisite precision the galaxy 2D 2PCF can be measured when significantly
larger data sets become available.
Fig.\ref{fig:Hzfg} (left panel) \cite{Chuang11} shows the averaged 2D 2PCF 
measured from 160 LasDamas mock catalogs compared with a theoretical model.
The contour levels are apparent in the measured 2D 2PCF even though no smoothing
is used (in contrast to the noisy current data, see Fig.\ref{fig:sdss_2d}); 
this is due to the reduction of shot noise achieved by averaging over
160 mock catalogs. Clearly, the 2D theoretical model used by \cite{Chuang11}
provides a reasonable fit to data on intermediate (and quasi-linear)
scales. 

A flux-limited galaxy redshift survey can
allow us to measure both $H(z)$ and $f_g(z)$ \cite{Guzzo07,Wang07}.
The measurement of $f_g(z)$ can be obtained through
independent measurements of redshift-space distortion parameter
$\beta=f_g(z)/b$ \cite{Kaiser87} and the bias
parameter $b(z)$ (which describes how light traces mass) \cite{Guzzo07}.
The parameter $\beta$ can be measured directly from galaxy redshift 
survey data by studying the observed redshift-space correlation function
\cite{2dFbeta,beta}. 
We can assume that the galaxy density perturbation $\delta_g$
is related to the matter density perturbation $\delta(\bfx)$
as follows \cite{Fry93}:
$\delta_g= b \delta(\bfx)+ b_2 \delta^2(\bfx)/2.$
The galaxy bispectrum is
$\langle \delta_{g\bfk_1} \delta_{g\bfk_2} \delta_{g\bfk_1}\rangle
= (2\pi)^3 \left\{P_g(\bfk_1) P_g(\bfk_2)\left[J(\bfk_1,\bfk_2)/b
+b_2/b^2\right] +cyc.\right\} \delta^D(\bfk_1+\bfk_2+\bfk_3),$
where $J$ is a function that depends on the shape of the
triangle formed by ($\bfk_1$, $\bfk_2$, $\bfk_3$) in
$\bfk$ space, but only depends very weakly on cosmology \cite{bias}.
Ref.\cite{bias} developed the method for measuring $b(z)$
from the galaxy bispectrum, which was applied by \cite{2dFbias}
to the 2dF data.
Independent measurements of $\beta(z)$ and $b(z)$
are very limited at present \cite{2dFbeta,2dFbias};
this will change dramatically in the near future.

The right panel of Fig.{\ref{fig:Hzfg}} shows how well a flux-limited
NIR galaxy redshift survey covering $>$10,000 square degrees and $0.5<z<2$ can 
constrain $H(z)$ and $f_g(z)$, compared with current data \cite{Wang07}.
The bottom half of the right panel in Fig.{\ref{fig:Hzfg}} shows the $f_g(z)$ 
for a modified gravity model (the DGP gravity model)
with $\Omega_m^0=0.25$ (solid line), as well as a dark energy model
that gives the same $H(z)$ for the same $\Omega_m^0$ (dashed line).
The cosmological constant model from the top half of the right panel
in Fig.{\ref{fig:Hzfg}} is also shown (dotted line).
Clearly, current data can not differentiate between dark energy
and modified gravity.
A very wide and deep galaxy redshift survey provides
measurement of $f_g(z)$ accurate to a few percent; this will allow an unambiguous distinction
between dark energy models and modified gravity models
that give identical $H(z)$ (see the solid and dashed lines
in the bottom half of the right panel of Fig.\ref{fig:Hzfg}).

The systematic effects of BAO as a standard ruler are: bias between
luminous matter and matter distributions, nonlinear effects, and 
redshift distortions \citep{bao_first}. Cosmological N-body
simulations are required to quantify these effects \cite{N-body}.
Ref.\cite{nonlinear1} shows that nonlinear effects can be accurately
taken into account. Ref.\cite{nonlinear2} shows that
the BAO signal is {\it boosted} when bias, nonlinear effects,
and redshift distortions are properly
included in the Hubble Volume simulation.
For a detailed discussion, see \cite{Wang_book}.

\section{Summary: Current Status and Future Prospects}

There are a large number of dark energy surveys that are ongoing, planned,
or proposed. Ongoing projects include Carnegie Supernova Project (CSP),
ESO Visible and Infrared Survey Telescope for Astronomy (VISTA) Surveys, 
Panoramic Survey Telescope \& Rapid Response System (Pan-STARRS), 
Hobby-Eberly Telescope Dark Energy Experiment (HETDEX), and 
Sloan Digital Sky Survey (SDSS) III, and the Dark Energy Survey (DES) \cite{ongoing}.
Selected future projects include Large Synoptic Survey Telescope (LSST) \cite{lsst},
and Euclid \cite{Euclid_RB,euclid}. Proposed future projects include 
the Wide-Field Infrared Survey Telescope (WFIRST), BigBOSS, and 
Square Kilometre Array (SKA) \cite{proposed}.
It is critical to remember that the challenge to solving 
the dark energy mystery will not be the statistics of the data obtained, 
but the tight control of systematic effects inherent in the data.
A combination of all three most promising methods (SNe Ia, GC, and WL)
should be used, each optimized by having its systematics minimized by design \cite{JEDI}.
It is an exciting time in cosmology. We can expect to make ground-breaking
discoveries about the nature of dark energy within the next decade or two.

I am grateful to Chris Blake, Daniel Eisenstein, Mario Hamuy, Dan Kasen, and Kevin Krisciunas
for permission to use their figures in this proceeding paper.
This work was supported in part by DOE grant DE-FG02-04ER41305.

\bibliographystyle{aipproc}   





\end{document}